# An Alignment Algorithm for Sequences

Sandeep Hosangadi and Subhash Kak

**Abstract.** This paper describes a new alignment algorithm for sequences that can be used for determination of deletions and substitutions. It provides several solutions out of which the best one can be chosen on the basis of minimization of gaps or other considerations. The algorithm does not use similarity tables and it performs aspects of both global and local alignment. The algorithm is compared with other sequence alignment algorithms.

## Introduction

Given a normative sequence and a fragment of a copy of it that in general is changed, we consider the problem of best alignment of the fragment. This alignment may be done under several assumptions regarding the sequences and it can provide functional, structural, or evolutionary relationships between the sequences. The alignment can be considered from a combinatorial perspective together with assumptions related to the nature of errors that can be deletions and substitutions.

The sequence characters in bioinformatics can be any genic (gene sequence, protein sequence), structural (morphological) or behavioral features of an organism. The DNA (deoxyribonucleic acid) sequence is a string of adenine (A), guanine (G), cytosine (C) and thiamine (T) bases and there are several approaches to the solution to the alignment problem of two DNA sequences. In general, the alignment we seek is not for all the bases of the fragment but rather of various parts of it at different locations. We may find different solutions where the parts of the fragment have different gaps associated with them. Given any number of sequences, finding out the degree of similarity between them is highly helpful in the field of bioinformatics, because similar regions help in deciding structural and evolutionary relationship between the sequences [1].

In database search by strings, the problem can be of errors in spellings or variant spellings or even systematic errors due to misalignment of hands with the keyboard. In such cases a similarity matrix between the characters needs to be considered. Likewise, there can be alternate spellings and representation of heard utterances in a variety of ways. In diachronic examination of and language (see, for example, [2]) the question of alignment must deal with changes. In biological memory, fragments may be matched by means of indices [3],[4] or by shape [5], which are aspects of alignment that go beyond string matching.

There exist various sequence alignment algorithms to find the best alignment between two sequences. In general these algorithms perform either global or local alignment or a combination of the two. Global alignment is generally performed if the sequence lengths are comparable whereas local alignment is when one is looking for the best fit within the larger sequence of the small sequence. The Needleman-Wunsch (NW) [6] and the Smith-Waterman (SW) [7] algorithms are well known algorithms for finding the best alignment between sequences. These algorithms make use of dynamic programming to find the best alignment and they use similarity tables. NW performs global optimization, whereas SW calculates local alignments for all possible length of sequences. For sequences of length *m* and *n*, the complexity of the two algorithms in their basic forms is *O(mn).* Accelerated versions of the algorithms run faster than this. For general review of alignment algorithms and distance measure, see [8],[9].

This paper presents an algorithm for aligning sequences that is based on matching subsequences with the determination of deletions and substitutions. The basic version of the algorithm does not use similarity tables, and it leaves gap or letter mismatch penalty as variables to be considered at the end. The complexity of the proposed algorithm is *O(mn).*

## Alignment Algorithms

We can find the best alignment between any two given sequences easily by brute force if their lengths are small. If the sequence lengths are large, we must develop a strategy to minimize the comparisons.

Basically, alignment methods perform global and local alignments. We will explain global alignment with an example consisting of two sequences S and V.

```
S:  ALKATKDSCKNSEBSEFDN
    |      ||||    |
V:  AHGFLERSCKNLMRLEDAH
```

The alignment is stretched over entire sequence lengths to match as many matches possible. Although, SCKN is the biggest match that is possible between S and V, A and E are also considered because of the match occurrence.

Local Alignment is an alignment that searches for segments of two sequences that match really well. Local alignment stops at the end of regions of similarity. It does not take the entire sequence into consideration. It just looks for regions that have more similar segments by neglecting regions with less similar segments.

Example for Local Alignment:

```
S: ----------SCKN----------
             ||||
V: ----------SCKN----------
```

## The Proposed Algorithm

We explain our algorithm by taking two sequences into consideration. We represent our first sequence as S and second sequence as V. Let the length of S be m and length of V be n, In general $n \leq m$. In our example, we take V to be shorter than S.

Our algorithm works as follows:

Step 1: Set K$\leftarrow$0, count2$\leftarrow$0 and count1$\leftarrow$0
Step 2: If $n-K \neq 1$, goto Step 3, Else goto Step 8

Step 3: Compare (n-k) size substring of T with (m-(n-K)) substring of S
Step 4: If match found, record the positions of V and S where match has occurred, Else goto Step 5.
Step 5: if (n-k) is less than n, Increment count2, else set count2←0 and goto Step 6.
Step 6: if (m-(n-k)) is less than m, Increment count1, else set count1←0 and goto Step 7
Step 7: Increment K, goto Step 2.
Step 8: Derive alignment string by printing similarity regions recorded when match occurred and fill the rest with gaps.

## Pseudocode:

```
ALIGNMENT(S, V)  // Alignment method with S and V
Begin
        m← S.length()
        n← V.length()
         Number_of_matches←0
        K←0
         Count2←0
         Count1←0
         While((n-k))>=1)
                  While((Count2+(n-k))<=n)
                           If(Count2+(n-k)<=n)
                                    Count1=0
                           While(Count1<m-(n-k))
                                  If((V.substring(Count2,Count2+(n-k))==S.substring(Count1,Count1+(n-
                                  k)))
                                            MARK[count1,count1+(n-k))
                                   // Record in S, where match has occurred.
                                            MARK[count2,count2+(n-k))
                                  // Record in V, where match has occurred.
                                  Count1++
                           Count2++
         Count1←0
         Count2←0
         k++
Alignment()  // Generates an alignment string by displaying all matches
             //and fills mismatches with gaps(-)
END
```

**Example 1**. Let S = FTFTALILLAVAV and V = FTALLAAV. The length m of S is 13 and the length n of V is 8 (m ≥ n). Since V is shorter, we perform comparison of sequence V with sequence S.

These are the comparisons that occur in the first iteration of our program:

FTALLAAV is compared with FTFTALIL, TFTALILL, FTALILLA, TALILLAV, ALILLAVA and LILLAVAV.

If match does not occur, K is incremented by one, so we perform (n-K) comparison of all possible strings in V with all possible (m-(n-k)) length strings in S.

FTALLAA and TALLAAV in sequence V are compared with FTFTALI, TFTALIL, FTALILL, TALILLA, ALILLAV, LILLAVA and ILLAVAV in sequence S

We perform such comparisons in sequence till (n-k) becomes 1 or all character in shorter sequence V gets match with sequence S.

Finally, we obtain alignment as:

**S:** F T F T A L I L L A V A V
**T:** - - F T A L - - L A – A V

**The problem of errors.** The sequences as presented to us may have errors. These errors could be of different kind: substitution errors, extra characters, dropped characters. Clearly, one must correct for errors based on knowledge of the database from where the strings have come. The matching process can point to likely places of substitution, extra characters, and dropped characters.

Let us assume we have an error in sequence S. Let the actual value of S = FTFTALILLAAV obtained after error correction. Now, if we perform alignment between S and V, we might get better alignment than what we got before. Likewise there could have been an error in the sequence V. If errors are known to have occurred it is important to correct these errors before using the alignment algorithm.

In the above example, if we consider S after error correction then we get a better result than that of the alignment that we got before.

$\mathbf{S_{new}}$ **=** F T F T A L I L L A A V
$\mathbf{T_{new}}$ **=** - - F T A L - - L A A V

$T_{new}$ has better alignment than T after considering error correction.

When there exists more than one match between any two sequences, we must select an alignment that has minimal gap amongst them or use other considerations related to similarity between characters.

Here we propose the following method to choose the better alignment: Generate all possible alignments between sequences and calculate the mean and variance for the gaps for each of those alignments. An alignment with less gap mean and smaller variance, if the means are the same, is the optimal alignment.

**Example 2.** Let us consider S= MNQRSETNLYATQRABCRTLJKYAMNTR and V= QRTLYATR. The possible alignments T that we could get are:

**S:** M N Q R S E NT L Y A T Q R A B C R T L J K Y A M N T R

**T:** - -  Q R - - - T L Y A T - R -  - - -  - - - - - - - - - - -

- -  Q R- - - T L Y A T - - - - - R - - - - - - - - - - -

- -  Q R - - -  T L Y A T - - - - - - - - - - - - - - - - R

In the above example, we have 3 possible alignments for T. The gap variance is computed after finding the mean of the gaps which are: 2, 4, and 9 respectively. The variance values for the gaps are: 1, 1, and 36 respectively. But still first one has minimum number of gaps, Therefore, the first alignment is the best one.

There are few words in English dictionary, which are frequently confused with words that sound same, but are spelled differently. For example effected and affected, to and too, sent and cent, here and hear and many more. These words are called as homonyms.

Same names are spelled in different ways in different regions or they are pronounced differently. For example Lata is an Indian name which is spelled as Latha in some parts of India. Likewise Sandeep is also spelt as Sandip and Vidya as Bidya. The English “a” has the pronunciation of “ai” in many words. If we are able to derive a matrix that tells us which letters could be substituted with existing letters that could minimize the gaps between sequences, and our algorithm could then work significantly better than existing ones. Example 2.1 illustrates the situation.

**Example 2.1**: Let S=MNQRSTLATATHSMNA and V = MNLATHA, The best possible alignment according to our algorithm would be MN----LA--TH---A. But if we consider homonyms into consideration, MN----LATA would be best alignment because the mean for the gaps is 4/3, whereas the previous one was 3. The value of variance without homonym consideration is 0.667, whereas with homonym consideration generates a variance of 0. Hence homonym considerations could help improve gap alignment in a significant way.

**Example 2.2:** Consider two false DNA sequences S and V, which are obtained from D sequences of two prime numbers [10],[11],[12]. We make use of 00 as A, 01 as G, 10 as C and 11 as T. Here are the following sequences that we obtained for two prime numbers 47 and 13.

**S =** A A G G G T A C G C A C A A C C T C G A T A G
**V** = A G A T C T

The possible alignments that would be generated according to our algorithm are :

**$T_1$** = - A G - - - A - - - - - - - - - - T C - - T
**$T_2$** = A - G - A- - - - - - T C - - T
**$T_3$**= - A G - - - - - - - - A - - - - - - T C - - T

We calculate mean and variance for $T_{1,}$ $T_2$ and $T_{3.}$ The means are 5.33, 2.5 and 5 respectively. Variances of $T_1$ $T_2$ and $T_3$ are 11.556, 2.25 and 4.667. Based on both mean and variance values, $T_2$ is the best alignment.

**Example 3:** Let us consider two biological DNA sequences S and V and find best alignment between them using our algorithm.
S=AGTCTAACTAGAATATACCGTACAGTACGAAG and T=TACTAGGAG, so that m= 32 and n=9. Here are the possible alignments our algorithm generates.

**S** = A G T C TA A C T A G A  A T A T A C C G T A C A G T A C G A A G

**$T_1$**= - - T - - - A C T A G - - - - - - - - G - - - A G

**$T_2$**= - - - - T - A C T A G - - - - - - - - G - - - - - - - - - - A G

**$T_{3=}$** - - - - T - A C T A G - - - - - - - - G - A - - G

**$T_{4=}$** - - - - T – A C T A G - - - - - - - - G - - - - - - A - G

There are many possible alignments that can be generated, Likewise, our algorithm generates all possible alignments and computes variance for all alignments. An alignment with less variance could be chosen as best alignment.

In the above example, mean of $T_{1,}$ $T_2$, $T_3$ and $T_4$ is 4.667, 6.33, 3 and 4 respectively. Variances are 5.556, 14.889, 8.5 and 9.5.

Since alignment string $T_1$ has minimum variance, $T_1$ is considered as best alignment amongst $T_{1,}$ $T_2$ $T_3$ and $T_{4.}$ This strategy defines a new way of selecting optimal alignment between given any number of sequences.

**Minimization of Gaps**

When we perform alignment between any sequences, our main objective is to minimize gaps between them. So we have to select optimal alignments by discarding all non-optimal alignments that would create more gaps. When given sequences are long, we need to use dynamic programming to calculate best possible alignment. It is also necessary to prune bad alignments in the beginning.

**Analysis of Complexity**

Given sequences S and V of length *m* and *n* respectively, the proposed algorithm performs $\sum_{k=0}^{n}(m-(n-k))*(n-k)$ comparisons to calculate alignment between them where *k* is an integer that ranges from 1 to *n*.

The complexity of our algorithm can be derived by the expansion of $\sum_{k=0}^{n}(m-(n-k))*(n-k)$

$$= (m-n)*n + ((m-n-1)*(n-1)) + ((m-n-2)*(n-2)) + \text{-----------} + mn$$
$$= mn+n^2 + mn-m-n^2-n-n+1+\text{-------------------------------------}+mn$$

Since the higher order terms are *m* and *n*, we get complexity of *O(mn)* after expansion.

**Comparison with NE and SW Algorithms.**
Let us consider two sequences S and V where S = AGTCTAACTAGAATATACCGTACAGTACGAAG and V = TACTAGGAG.

1. Alignment using Needleman Wunsch Algorithm:

```
AGTCTAACTAGAATACCGTACAGTACGAAG
  || |||| |                |
  TA-CTAGGA----------------G
```

2. Alignment using Smith Waterman Algorithm:

```
AGTCTAACTAGAATACCGTACAGTACGAAG
           || |||| |
----TA-CTAGGAG----------------
```

3. Alignment generated by our algorithm:

```
AGTCTAACTAGAATACCGTACAGTACGAAG
      | |||||      | | ||
    ----T-ACTAG------G-A—AG
```

**Conclusions**
This paper describes a new alignment algorithm for sequences that can be used for determination of deletions and substitutions. Insertions may also be handled if the order of the comparison sequences is switched but, of course, it would make sense only if the size of the two sequences is comparable. Our algorithm provides several solutions out of which the best one can be chosen on the basis of minimization of gaps or other considerations. Statistical consideration related to alignment solutions can be studied in a manner similar to those for other alignment algorithms [13]-[17].

The proposed algorithm generates good alignment by finding maximum length matches between given sequences without having to use similarity tables and its complexity is of the same order as the MW and SW algorithms. The basic algorithm can be refined by adding further constraints related to character similarity or gap constraint that will also make it more efficient since that would exclude many bad solutions.